\documentclass[runningheads]{llncs}

\newif\ifEditMode
\EditModefalse

\newif\ifAddProofs
\AddProofstrue

\usepackage{graphicx, amsmath, amsfonts,amssymb, tikz, mathtools, tikz-cd,epstopdf,verbatim, hyperref}
\usepackage{wrapfig,xspace,subcaption}
\usepackage[font=itshape]{quoting}
\usepackage{enumitem}
\usepackage{amsbsy,scalerel}
\usepackage[T1]{fontenc}
\usepackage{algorithm}
\usepackage[noend]{algpseudocode}
\usepackage{syntax}
\usepackage[colorinlistoftodos]{todonotes}
\usetikzlibrary{automata, positioning, arrows}

\usepackage{amsthm}

\tikzset{
->, 
node distance=3cm, 
every state/.style={thick, fill=gray!10}, 
initial text=$ $, 
}

\setlist[enumerate]{align=right,itemindent=2em,labelsep=2pt,labelwidth=1em,leftmargin=0pt,nosep}

\PassOptionsToPackage{bookmarks=false}{hyperref}

\graphicspath{{figures/}}

\allowdisplaybreaks

\newcommand{\iir}[1]{\ifEditMode\todo[size=\footnotesize,color=blue!40]{iir: #1}\fi}

\ifEditMode
   \paperwidth=\dimexpr \paperwidth + 8cm\relax
   \oddsidemargin=\dimexpr\oddsidemargin + 4cm\relax
   \evensidemargin=\dimexpr\evensidemargin + 4cm\relax
   \marginparwidth=\dimexpr \marginparwidth + 4cm\relax
\fi

\DeclareCaptionFormat{algor}{%
  \hrulefill\par\offinterlineskip\vskip1pt%
    \textbf{#1#2}#3\offinterlineskip\hrulefill}
\DeclareCaptionStyle{algori}{singlelinecheck=off,format=algor,labelsep=space}
\usepackage{xcolor}

\newcommand{\setunion}[0]{\boldsymbol{\cup}}
\newcommand{\setint}[0]{\boldsymbol{\cap}}

\newcommand{\behaviorUnv}[0]{\mathcal{B}}

\newcommand{\satimpl}[0]{\models^{I}}
\newcommand{\satenv}[0]{\models^{E}}

\newcommand{\cont}{\mathcal{C}}

\renewcommand{\paragraph}[1]{\noindent\textbf{#1}}

\renewcommand{\tt}{\texttt}

\newcommand{\reals}[0]{\mathbb{R}}
\newcommand{\realsnn}[0]{{\reals_{\ge 0}}}
\newcommand{\naturals}[0]{\mathbb{N}}
\newcommand{\varset}[0]{\mathbf{VarSet}}
\newcommand{\den}[0]{\mathbf{Den}}

\newcommand{\tlist}[0]{\mathfrak{t}}
\newcommand{\alist}[0]{\mathfrak{a}}
\newcommand{\glist}[0]{\mathfrak{g}}

\newcommand{\transp}[0]{\intercal}

\newcommand{\pacti}{Pacti\xspace}

\newcommand{\qeda}{\hfill\rule{1ex}{1ex}}

\begin{document}
\title{\pacti: Scaling Assume-Guarantee Reasoning for System Analysis and Design\thanks{The theory in this paper is based on Chapter 7 of \cite{Incer:EECS-2022-99}.}
}

\author{Inigo Incer\inst{1,3}
\and
Apurva Badithela\inst{3}
\and
Josefine Graebener\inst{3}
\and
Piergiuseppe Mallozzi\inst{1}
\and
Ayush Pandey\inst{3}
\and
Sheng-Jung Yu\inst{1}
\and
Albert Benveniste\inst{2}
\and
Benoit Caillaud\inst{2}
\and
Richard M. Murray\inst{3}
\and
Alberto Sangiovanni-Vincentelli\inst{1}
\and
Sanjit A. Seshia\inst{1}
}
\authorrunning{Incer et al.}
\institute{University of California, Berkeley, USA
\and
INRIA/IRISA, Rennes, France
\and
California Institute of Technology, USA
}

\maketitle
Contract-based design is a method to facilitate modular system design. While there has been substantial progress on the theory of contracts, there has been less progress on scalable algorithms  for the algebraic operations in this theory. In this paper, we present: 1) principles to implement a contract-based design tool at scale and 2) \pacti, a tool that can efficiently compute these operations. We then illustrate the use of \pacti in a variety of case studies.

\section{Introduction}
\vspace{-5pt}
It has been fifteen years since contract-based design \cite{Benveniste2008,BenvenisteContractBook} was proposed as a formal methodology to facilitate the design of general cyber-physical systems. A key idea is to represent components in a system using assume-guarantee specifications, or \emph{contracts}. Contracts aim to enable two processes: independent design and concurrent design. By independent design, we mean that a set of contracts whose composition refines a top-level requirement is identified. These new contracts can be delivered to others in order to obtain an implementation, all while knowing in advance that the composition of these implementations will meet the top-level objective. By concurrent design, it is meant that the specification of a given subsystem can be broken into multiple specifications, each addressing a certain viewpoint (e.g., functionality or performance) of the design element in question. Analysis can then be carried out by using as much specification data as needed for the task at hand.

A rich contract algebra has been developed to carry out operations of relevance to system analysis and design. Using this algebra, we can compare contracts to discern when a specification is more \emph{refined}, or stricter, than another. Through the operation of \emph{composition}, we can compute the specification of the system built by composing implementations of several specifications. Given a top-level specification and given the specification of a subsystem that will be used to build the system, the \emph{quotient} is the specification of a missing subsystem whose composition with the existing subsystem yields a system meeting the desired top-level specification.

Given its importance in the system-design process, tools to support contract-based design have been developed.
OCRA~\cite{Cimatti13} supports refinement checking of temporal contracts;
AGREE~\cite{Cofer12} uses assume-guarantee reasoning to hierarchically verify contract refinement using past-time linear temporal logic;
CHASE~\cite{NuzzoDATE18} combines front-end formal specification language with back-end requirement correctness, completeness, and refinement checking;
CROME~\cite{mallozzi2020crome} allows engineers to refine and realize robotic-mission specifications using contracts.
These tools use the contract-based methodology to verify systems hierarchically
but offer limited support to compute system-level contracts and missing-component contracts. In particular, 
they do not enforce contracts to be expressed using only the interfaces of the components, limiting the scalability of contract operations and their use in compositional design.

This paper presents the theoretical foundations and design of \pacti, a tool that allows designers to efficiently manipulate contracts.
The contracts generated by \pacti
are defined only using the interfaces of the components, which increases their readability by human designers and reduces their complexity for further processing by tools.
The structure of the paper is as follows: Section \ref{sc:agIntro} provides a brief overview of assume-guarantee contracts and introduces a formalism to deal with syntactic representations of contracts. Section \ref{sc:principles} introduces the three principles that allow us to compute contracts for complex systems and that form the basis of \pacti. Section \ref{contractoperations} describes the algorithms used to efficiently compute the contract operations.
Section \ref{sc:casestudies} discusses applications of \pacti to problems in autonomous systems, multi-agent systems, synthetic biology, and digital circuit design.

\vspace{-3mm}
\section{Assume-guarantee contracts}
\vspace{-2mm}
\label{sc:agIntro}

We provide the standard definitions of assume-guarantee contracts and then introduce a formalization for connecting contracts to syntactic representations.

\medskip
\paragraph{Background on assume-guarantee contracts.}
We follow \cite{BenvenisteContractBook} in our definitions.
Suppose a universe $\behaviorUnv$ of behaviors is given that can include behaviors over system variables (cyber, physical, functional, nonfunctional, etc.). 
The components over which we define predicates and the predicates themselves are
defined as sets of behaviors. We define a component $M \subseteq
\behaviorUnv$ as the collection of behaviors we can witness from it, whereas a
property $P \subseteq \behaviorUnv$ is the set of behaviors meeting
a given criterion, such as safety.
We say a component $M$ satisfies a property $P$, written $M \models P$, if $M \subseteq P$. Given two properties $P$ and $P'$, we say that $P$ is a refinement of $P'$ if for every component $M$, $M \models P \Rightarrow M \models P'$. Thus, $P$ is a refinement of $P'$ if $P \le P'$, where $\le$ is the subset order.
Component composition is given by set intersection, i.e., for two components $M, M'$, their composition is $M \parallel M' = M \setint M'$.

A contract is a pair of properties $\cont = (A, G)$, where $A$ represents the \emph{assumptions}, and $G$ the \emph{guarantees} of the contract.
A component $E$ is called an \emph{environment} of the contract, denoted $E \satenv \cont$ if $E \models A$. A component $M$ is an implementation of the contract, denoted $M \satimpl \cont$, if $M \parallel E \models G$ for all $E \satenv \cont$. That is, a component is an implementation of a contract if it satisfies the contract's guarantees when operating in an environment that meets the contract's assumptions. We discuss the order and various operations of contracts.

\begin{enumerate}
\item \emph{Refinement}.
Contracts are partially ordered \cite{BenvenisteContractBook}. Given contracts $\cont = (A, G)$ and $\cont' = (A', G')$, we say that $\cont$ is a \emph{refinement} of $\cont'$ (or that $\cont'$ is a relaxation of $\cont$), denoted $\cont \le \cont'$, if any implementation of $\cont$ is an implementation of $\cont'$ and any environment of $\cont'$ is an environment of $\cont$:
$\cont \le \cont' \Leftrightarrow (A' \le A) \land ( G \setunion \neg A \le G' \setunion \neg A' )$.
This order generates a well-defined greatest-lower bound, called \emph{conjunction}. It is given by $\cont \land \cont' = \big( (A \setunion A'), (G \setunion \neg A) \setint (G' \setunion \neg A') \big)$. Conjunction yields a contract that retains all information about the contracts being conjoined: the guarantees of the contracts being conjoined are required to hold when their corresponding assumptions hold. This operation is used to merge viewpoints that do not need to hold simultaneously.

\item \emph{Composition}.
The operation of composition \cite{BenvenisteContractBook} allows us to obtain a specification for a system built by composing implementations of the contracts being composed. Its closed-form expression is
$\cont \parallel \cont' = \big( (A \setint A') \setunion (A \setint\neg G) \setunion (A' \setint \neg G') , (G \setunion \neg A) \setint (G' \setunion \neg A') \big)$.

\item \emph{Quotient}.
The quotient \cite{agquotient} allows us to solve the following problem: given a top level specification $\cont$ that we want a system to meet and given the specification $\cont'$ of a partial implementation of the system, what is the specification of the component that we are missing to implement $\cont$? We can compute it as follows:
$\cont / \cont' = \left( A \setint (\neg A' \setunion G') , (A' \setint G) \setunion \neg A  \setunion (A' \setint \neg G') \right)$.

\item \emph{Merging}
(or strong merging) \cite{agContractMerging} can be used to 
handle multiple contract viewpoints that need to be enforced simultaneously. This operation yields a contract whose environments satisfy the assumptions of both contracts and whose implementations are valid implementations for both contracts:
$\cont \bullet \cont' = \left( A \setint A', (G \setint G') \setunion \neg A \setunion \neg A' \right)$.
\end{enumerate}

\medskip
\paragraph{Syntax.}
The standard definitions of assume-guarantee contracts do not lend themselves to implementations. We need syntactic representations of contracts for this purpose. Here we establish the formal aspects of these representations.

Contracts are defined over a universe of behaviors, $\behaviorUnv$. We consider the construction of such a universe. The most fundamental concept in modeling is the variable. Variables are names associated with a concept in our system. We will define a variable as a tuple $(V, \behaviorUnv_V)$ of a name $V$ and a set of behaviors $\behaviorUnv_V$ for that variable. For instance, a static variable with real values would have $\behaviorUnv_V = \reals$, while a signal with discrete transitions and values taken in a domain $D$ would have a universe $\naturals \to D$. Continuous functions are defined by changing the domain: $\realsnn \to D$. Assume we have a set $\varset$ of variables in our system. Then we can build the universe of behaviors as $\behaviorUnv = \prod_{V \in \varset} \behaviorUnv_V$.

We assume we have access to a Boolean algebra $T$, called the \emph{term algebra}, whose elements we will call \emph{terms} or \emph{constraints}. $T$ will give us the syntax we will use to represent specifications on top of which contracts are built.
This Boolean algebra comes equipped with a Boolean map $\den\colon T \to 2^\behaviorUnv$ called the \emph{denotation map}. The fact $\den$ is a Boolean map means it commutes with the Boolean algebra structure of $T$.
\iir{Define what it means for a term to ``depend'' on a variable.}

For $a, g \in T$, we can write contracts over $T$ as $\cont = (a, g)$, where $a$ and $g$ are terms. Applying the denotation map element-wise enables us to connect contracts over $T$ with contracts over $2^\behaviorUnv$: $\den\colon (a,g) \mapsto (\den(a), \den(g))$.
We can compute the contract operations \emph{in the term algebra} as follows: for contracts $\cont = (a,g)$ and $\cont' = (a', g')$ defined over the term algebra, using the operations stated in the background discussion, we have
{\small
\begin{align}\label{eq:refinement}
\cont \le \cont' &\Leftrightarrow (a' \le a) \land ( g \lor \neg a \le g' \lor \neg a' )
\\
\label{eq:composition}
\cont \parallel \cont' &= \left( (a \land a') \lor (a \land\neg g) \lor (a' \land \neg g') , (g \lor \neg a) \land (g' \lor \neg a') \right)
\\
\label{eq:quotient}
\cont / \cont' &= \left( a \land (\neg a' \lor g') , (a' \land g) \lor \neg a  \lor (a' \land \neg g') \right)
\\
\label{eq:merging}
\cont \bullet \cont' &= \left( a \land a', (g \land g') \lor \neg a \lor \neg a' \right)
\end{align}}%

\vspace{-5mm}
\section{Reducing complexity of contract operations}
\label{sc:principles}

\vspace{-2mm}
The definitions of refinement~\eqref{eq:refinement}, composition~\eqref{eq:composition}, quotient~\eqref{eq:quotient}, and merging~\eqref{eq:merging} over a term algebra $T$ immediately suggest a difficulty: the results of these system construction/deconstruction operations are considerably more complex than the original specifications themselves. This brings problems to 1) the generation of specifications that a designer can readily understand and to 2) the further automated processing of specifications, as now the algorithms have to manipulate longer formulas.

Our objective is to identify means for reducing the complexity of the computed operations to make them more understandable to designers, and more succinct, so that contract operations can be applied repeatedly without the constraints exploding in length. We adopt three principles towards this end. In this section, we discuss these principles used to implement \pacti and present algorithms for the efficient computation of some contract operations.


\paragraph{Principle 1: post-processing contract operations.} All contract operations are defined as a contract satisfying certain optimality criteria. For example, composition is defined as the smallest contract such that the composition of the implementations of the contracts being composed satisfies the guarantees of the contract and the environments of the contract satisfy other criteria (see \cite{BenvenisteContractBook}, Chapter 4 or \cite{Incer:EECS-2022-99}, Chapter 6). Similarly, given contracts $\cont$ and $\cont'$, the quotient is defined as the largest contract $\cont''$ such that $\cont' \parallel \cont'' \le \cont$, i.e., the quotient is the largest specification of a missing component that will allow a system to meet a top-level specification. The key observation is that \emph{the fact that contract operations are optimal does not mean they should be the end result we should communicate to the user or keep for subsequent processing by tools.} For example, a system that obeys the composite specification will also obey a looser (i.e., more relaxed) contract. Similarly, any specification that refines the quotient will also be the specification of a missing component. We conclude that
\emph{if an operation is defined as a minimum (resp. maximum), then a relaxation (resp. refinement) of the operation can be returned to the user.} Thus, we will relax or refine the contract operations in order to place contracts in a more desirable form. The next two principles address the form that contracts should have and that will yield an algorithm for contract post-processing.

\smallskip
\paragraph{Principle 2: contracts as lists of requirements.} Requirements in industry are often expressed as conjunctions of constraints. In general, component datasheets state a list of requirements that must hold simultaneously on the environment (e.g., bounded temperature, input voltages, etc.) in order for a list of promises to hold.
We will call \emph{termlist} the elements of $2^T$. The denotation of termlists will be given by the composed map $\begin{tikzcd}[column sep=small]2^T \arrow[r, "\land"] & T \arrow[r, "\den"] & 2^\behaviorUnv \end{tikzcd}$ which works as follows: $\tlist  \mapsto \den (\land_{t \in \tlist} t) = \setint_{t \in \tlist} \den(t)$. Therefore, \emph{we will express contracts in the form $(\alist, \glist)$, where $\alist,\glist \in 2^T$, indicating that several promises hold as long as several assumptions hold}.

\smallskip
\paragraph{Principle 3: IO contracts.} Influenced by IO Automata \cite{Lynch89anintroduction}, Interface Automata \cite{10.1145/503271.503226}, and Moore Interfaces \cite{mooreInterfaces}, we will extend the definition of a contract to be
\begin{definition}
Let $T$ be a term algebra and $\varset$ be a set of variables. An IO contract is a list $(I, O, \alist, \glist)$, where $I, O \subseteq \varset$ are disjoint sets of input and output variables, respectively, and $\alist, \glist \in 2^T$ are lists of terms representing the assumptions and guarantees of the contracts, respectively. The assumptions of IO contracts only depend on input variables, and the guarantees only depend on input and output variables.
\end{definition}
The IO profile of a contract will play a key role when computing the contract operations, as discussed below.

\iir{Explain that term and termsets have a partial order tied to the denotation of the formulas.}

\vspace{-10pt}
\section{Implementing the contract operations}
\label{contractoperations}
\vspace{-10pt}
Our objective is to devise algorithms to compute the contract operations taking as inputs IO contracts and producing as outputs IO contracts. First we focus on composition. Consider two IO contracts $\cont = (I, O, \alist, \glist)$ and $\cont' = (I', O', \alist', \glist')$. Their composition will only be defined when $O$ and $O'$ are disjoint. When this happens, we have $\cont \parallel \cont' = (I_c, O_c, \alist_c, \glist_c)$, where $I_c = (I \setunion I') \setminus (O \setunion O')$ and $O_c = (O \setunion O') \setminus (I \setunion I')$. This operation will not keep in the composed contract's IO profile any output which serves as an input of the other contract being composed; this means that, in general, composition of IO contracts is not associative.\footnote{In the future, we plan to explore a composition operation that allows the user to specify which of the output signals should be kept in the composed contract.}

$\alist_c$ and $\glist_c$ are computed as follows. Let $a = \land \alist$, $g = \land \glist$, $a' = \land \alist'$, and $g' = \land \glist'$. We form the $T$-contract $(a_c, g_c)$ for $a_c, g_c \in T$, where the contract $(a_c, g_c)$ is given by the standard contract composition \eqref{eq:composition}.
Since $\cont$ and $\cont'$ are IO contracts, $a$ and $g$ have terms depending on $I$ and $I \setunion O$, respectively, and similarly for $a'$ and $g'$. This means that $a_c$ and $g_c$ will fail to produce an IO contract because the assumptions and guarantees will depend on variables other than those allowed (i.e., $I_c$ for the assumptions and $I_c \setunion O_c$ for the guarantees).
In order to produce an IO contract after composition, we apply Principle 1. Since we know that composition is defined as a minimum, we can relax the operation \eqref{eq:composition} in order to obtain a well-defined IO contract. How should such a relaxation be computed? Per \eqref{eq:refinement}, to relax the contract means to refine the assumptions and loosen the guarantees. We observe that the assumptions of the composition \eqref{eq:composition} have three terms: $a \land a'$, $a \land\neg g$, and $a' \land \neg g'$. We refer to the first term as the \emph{stem} of the assumptions since this term represents the simultaneous enforcement of the constraints of $a$ and $a'$ and thus means that we can expect the guarantees of both contracts to hold. The terms $a \land \neg g$ and $a' \land \neg g'$ are \emph{failure terms}, as they state that the assumptions of a contract were met, but the component did not deliver its promises. These terms are used to carry out transformations in the stem, as shown in the theorem below, but we can safely remove them after we have used them. \emph{We transform the stem in order to remove from it references to variables that do not belong to $I_c$}.
\begin{theorem}\label{th:kbzjhgj}
    Let $(a, g)$ and $(a',g')$ be contracts defined over a term algebra $T$. Let $\cont_c = (a_c, g_c)$ be the composition, as computed by \eqref{eq:composition}. Suppose that $a'' \in T$ satisfies $a'' \land g \land a \le a' \land g \land a$ and $g'' \in T$ satisfies $g\land g' \le g''$. Then the contract $(a\land a'', g'')$ is a relaxation of $\cont_c$.
\end{theorem}
\ifAddProofs
\begin{proof}
    We compute
    $
        a \land a'' 
        \le (a \land a'') \lor (a \land \neg g) \lor (a' \lor \neg g') 
        = (a \land a'' \land g) \lor (a \land \neg g) \lor (a' \lor \neg g') 
        \le (a \land a' \land g) \lor (a \land \neg g) \lor (a' \lor \neg g')
        = (a \land a') \lor (a \land \neg g) \lor (a' \lor \neg g').
    $
    Thus, $a \land a'' \le a_c$. We also have
    $
        g'' \lor \neg (a \land a'') 
        \ge (g \land g') \lor \neg a \lor \neg a''
        \ge (g \land g') \lor \neg a \lor (a \land g \land \neg a'') 
        \ge (g \land g') \lor \neg a \lor (a \land g \land \neg a')
        = (g \land g') \lor \neg a \lor (g \land \neg a')
        \ge (g \land g') \lor (g' \land \neg a) \lor (\neg a \land \neg a') \lor (g \land \neg a') = g_c \lor \neg a_c.
    $
    We conclude that the contract $(a \land a'', g'')$ is a relaxation of $\cont_c$.
\end{proof}
\fi
If we have IO contracts $\cont = (I, O, \alist, \glist)$ and $\cont' = (I', O', \alist', \glist')$ and their composition is defined (the sets of output variables are disjoint), we use Theorem \ref{th:kbzjhgj} to compute their composition $(I_c, O_c, \alist \setunion \alist'', \glist'')$. This means we have to identify termlists $\alist''$ and $\glist''$ such that $(\land \alist'') \land (\land \glist) \land (\land \alist)\le (\land \alist') \land (\land \glist) \land (\land \alist)$ and
$\land \glist'' \ge (\land \glist) \land (\land \glist')$. Using proof-theoretic notation, we can consider the sets of constraints $\glist \setunion \alist$ as a context for the following inference: $\glist, \alist, \alist'' \vdash \alist'$; i.e., we use the context $\glist \setunion \alist$ in order to refine the terms $\alist''$ from $\alist'$. Similarly, we use the terms $\alist \setunion \alist''$ to relax $\glist''$ from $\glist \setunion \glist'$. Observe that Theorem \ref{th:kbzjhgj} only allows us to refine $\alist'$ using the context $\glist$ or $\alist$ using the context $\glist'$. We pick the context based on the interconnection of the contracts. If a component drives the inputs of another component, and the assumptions of the second component depend on this driven input,
we use the former's guarantees as a context to refine the latter's assumptions. From our considerations so far, we will not allow both components to have outputs driving each other's inputs when the assumptions of both components depend on those inputs. Procedure \textproc{ContractComposition} of Algorithm \ref{nlkqbl} shows how we compute IO-contract composition.

\textproc{RefineWithContext}$(\tlist, \tlist', S)$ and \textproc{RelaxWithContext}$(\tlist, \tlist', S)$ refine and relax, respectively, a termlist $\tlist$ using the context $\tlist'$. The resulting termlist is only allowed to contain terms referring to the variables contained in the set $S$. The function $\textproc{Reduce}(\tlist, \tlist')$ is used to eliminate from $\tlist$ any redundant constraints, using the context $\tlist'$. 
The function $\textproc{IsRefinement}(\tlist, \tlist')$ tells whether the satisfaction of $\tlist$ implies the satisfaction of $\tlist'$. 
The implementations of these four functions depend on the specification theory in which the terms are expressed. 

\noindent
{\bf {Polyhedral constraints.}} This is the first theory supported by \pacti. The terms are linear inequalities with real coefficients.
To implement \textproc{RefineWithContext} and \textproc{RelaxWithContext},
we eliminate variables from a term by computing refinements and relaxations of linear inequalities using the algorithms presented in~\cite{varElimTechRp}. To implement \textproc{Reduce}, we use standard methods for the elimination of redundant terms, e.g., \cite{10.1007/978-3-319-52234-0-20,10.1287/mnsc.29.10.1209}.
To compute \textproc{IsRefinement}, i.e., to
verify whether a polyhedron is contained inside another, 
one can use linear programming, as shown by \cite{eaves1982optimal} in their solution of the ``HH formulation'' of the optimal containment problem.

\algnewcommand{\LeftComment}[1]{\(\triangleright\) #1}
{\footnotesize
    \captionsetup{hypcap=false}
    \captionof{algorithm}{Contract operations}\label{nlkqbl}
    \begin{algorithmic}[1]
        \renewcommand{\algorithmicwhile}{\textbf{procedure}}
        \newcommand{\algorithmicendwhile}{\algorithmicend\ \algorithmicwhile}
        \Statex
        \textbf{Input:} IO contracts $(I, O, \alist, \glist)$ and $(I', O', \alist', \glist')$
        \Statex
        \textbf{Output:} Relaxation of the composition operation
        \While {$\textproc{ContractComposition}(I, O, \alist, \glist, I', O', \alist', \glist')$}
        \State $O^c \gets (O \setunion O') \setminus (I \setunion I')$ \label{jcgadj}
    \State $I^c \gets (I \setunion I') \setminus (O \setunion O')$ \label{kjgsaghx}
    \State $I_c, I_c' \gets \text{Vars}(\alist), \text{Vars}(\alist')$ \Comment{Constrained inputs}
    \State $\text{CyclePresent} \gets (O' \setint I \ne \emptyset \textbf{ and } O \setint I' \ne \emptyset)$
    \If{$O \setint O' \ne \emptyset$ \textbf{or} (CyclePresent \textbf{and} $(O' \setint I_c \ne \emptyset$ \textbf{or} $O \setint I_c' \ne \emptyset)$)}
        \State \Return Error: Contracts are not composable
    \ElsIf{$O' \setint I \ne \emptyset$ and $O \setint I' = \emptyset$}
        \Statex \quad \quad \; \; \LeftComment{Refining $\alist$ using $\glist'$ and $\alist'$}
        \State $\bar \alist \gets \textproc{RefineWithContext}(\alist, \alist' \setunion \glist', I^c)$
        \State $\alist^c \gets \textproc{Reduce}(\bar \alist \setunion \alist',\emptyset)$ 
    \ElsIf{$O' \setint I = \emptyset$ and $O \setint I' \ne \emptyset$} \label{kjhdgkj}
        \Statex \quad \quad \; \; \LeftComment{Refining $\alist'$ using $\glist$ and $\alist$}
        \State $\bar \alist' \gets \textproc{RefineWithContext}(\alist', \alist \setunion \glist, I^c)$\label{kaglajk}
        \State $\alist^c\gets \textproc{Reduce}(\bar \alist' \setunion \alist,\emptyset)$\label{lqndhlq}
    \ElsIf{($O' \setint I_c = \emptyset$ \textbf{and} $O \setint I_c' = \emptyset$) \textbf{or} CyclePresent}
        \State $\alist^c\gets \textproc{Reduce}(\alist \setunion \alist', \emptyset)$
    \EndIf
    \Statex \quad \; \LeftComment{Find $\glist^c$ such that $(\land \glist^c) \land (\land \alist^c) \ge (\land \glist) \land (\land \glist') \land (\land \alist^c)$}
    \State $\glist^c \gets \textproc{RelaxWithContext}(\glist \setunion \glist', \alist^c, I^c \setunion O^c)$\label{lqnufgkqw}
    \State \Return $(I^c, O^c, \alist^c, \textproc{Reduce}(\glist^c,\alist^c))$
        \EndWhile
    \end{algorithmic}
\begin{algorithmic}[1]
    \renewcommand{\algorithmicwhile}{\textbf{procedure}}
    \newcommand{\algorithmicendwhile}{\algorithmicend\ \algorithmicwhile}
    \Statex \textbf{Input:} IO contracts $\cont = (I, O, \alist, \glist)$ and $\cont' = (I', O', \alist', \glist')$
    \Statex \textbf{Output:} Refinement of the quotient $\cont / \cont'$
    \While {$\textproc{ContractQuotient}(I, O, \alist, \glist, I', O', \alist', \glist')$}
\If{$I \setint O' \ne \emptyset$}
    \State \Return Error: The quotient is not defined
\EndIf
\State $O^q = (O \setminus O') \setunion (I' \setminus I)$\label{qkgsdj}
\State $I^q = (O' \setminus O) \setunion (I \setminus I')$\label{kbcg}
\State $\alist'' \gets \alist$\label{qkgcqjhg}
\If{$\textproc{IsRefinement}(\alist, \alist')$}
\State $\alist'' \gets \textproc{Reduce}(\alist'' \setunion \glist', \emptyset)$
\EndIf
\State $\alist'' \gets \textproc{RelaxWithContext}(\alist'', \emptyset, I^q)$\label{lqhlq}
\State $\glist'' \gets \textproc{RefineWithContext}(\glist, \alist' \setunion \glist', I^q \setunion O^q)$\label{lkjxgsdb}
\State $\glist''' \gets \textproc{RefineWithContext}(\alist' \setunion \glist'', \alist, I^q \setunion O^q)$\label{lkjdan}
\State \Return $(I^q, O^q, \alist'', \textproc{Reduce}(\glist''',\alist''))$
\EndWhile
\end{algorithmic}
\begin{algorithmic}[1]
    \renewcommand{\algorithmicwhile}{\textbf{procedure}}
    \newcommand{\algorithmicendwhile}{\algorithmicend\ \algorithmicwhile}
    \Statex \textbf{Input:} IO contracts $(I, O, \alist, \glist)$ and $(I', O', \alist', \glist')$
    \Statex \textbf{Output:} Strong merging
    \While {$\textproc{ContractMerging}(I, O, \alist, \glist, I', O', \alist', \glist')$}
\If{$(I \ne I') \textbf{ or } (O \ne O')$}
    \State \Return Error: Merging is not defined
\EndIf
\State $a^m \gets \textproc{Reduce}(\alist \setunion \alist',\emptyset)$
\State \Return $(I, O, a^m, \textproc{Reduce}(\glist \setunion \glist',a^m))$
\EndWhile
\end{algorithmic}
} 

\begin{figure}[t]
\centering
\begin{subfigure}[b]{0.35\textwidth}
\centering
\includegraphics[width=0.9\textwidth]{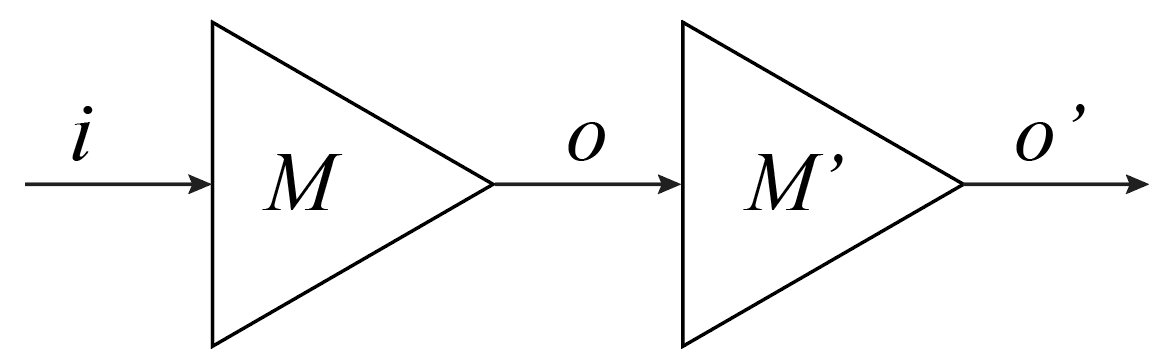}
\caption{}
\label{fg:comp-example}
\end{subfigure}
\begin{subfigure}[b]{0.35\textwidth}
\centering
\includegraphics[width=0.9\textwidth]{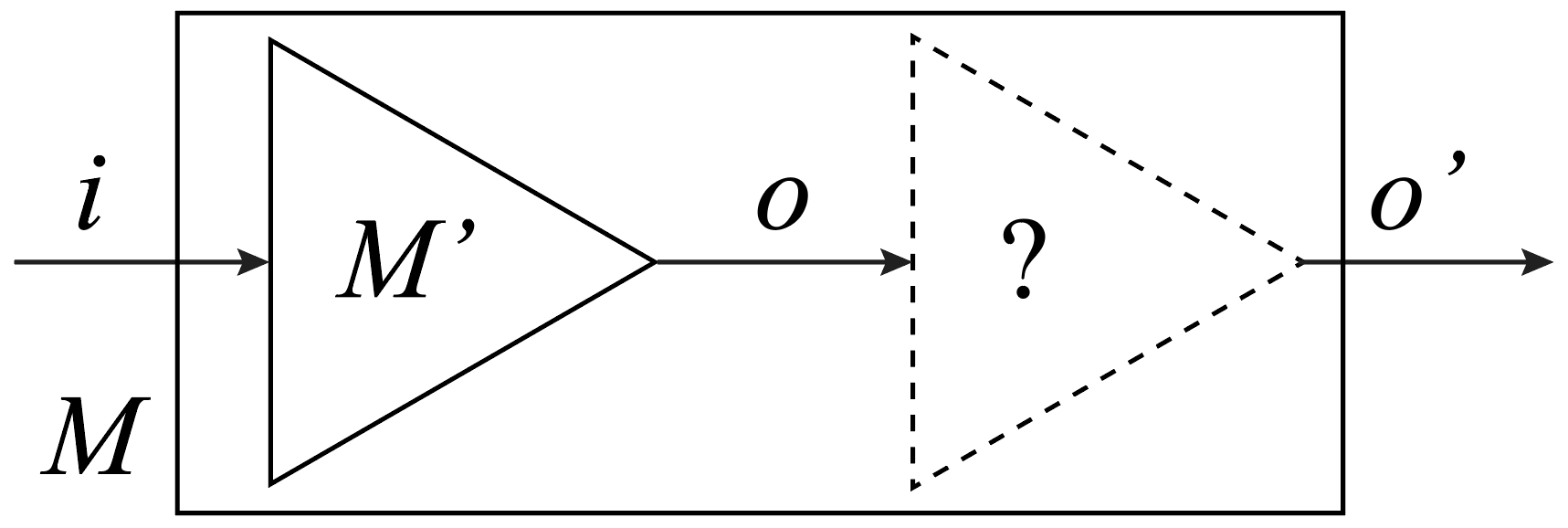}
\caption{}
\label{fg:comp-quotient}
\end{subfigure}
\vspace{1mm}
\caption{
\footnotesize
(a) Given the contracts for two buffers connected in series, we wish to compute the system-level contract. (b) Given
a top-level contract and the contract for one of the subsystems, we wish to compute the contract of the missing component.}
\end{figure}

\begin{example}
Consider the circuit of Figure~\ref{fg:comp-example}. We have components $M$ and $M'$ obeying contracts
$\cont = (\{i\}, \{o\}, \{i \le 2\}, \{o \le i\})$ and 
$\cont = (\{o\}, \{o'\}, \{o \le 1\}, \{o \le o\})$, respectively. We use \textproc{ContractCompose} of Algorithm~\ref{nlkqbl} to obtain the system-level contract $(I^c, O^c, \alist^c, \glist^c)$. L\ref{jcgadj}-L\ref{kjgsaghx} yield $I^c = \{i\}$ and $O^c = \{o'\}$. Condition~L\ref{kjhdgkj} is active since $M$ drives its outputs to the inputs of $M'$ (and not vice versa).
L\ref{kaglajk} yields
$\bar \alist' = \textproc{RefineWithContext}(\{o \le 1\}, \{i \le 2, o \le i\}, \{i\}) = \{i \le 1\}$,
and from L\ref{lqndhlq}, we obtain
$\alist^c = \{i \le 1\}$.
Finally, from L\ref{lqnufgkqw}, we get
$\glist^c = \textproc{RelaxWithContext}(\{o \le i, o' \le o\}, \{i \le 1\}, \{i,o'\}) = \{o' \le i\}$.
The resulting specification uses exclusively the inputs and outputs of the top-level system.
\qeda
\end{example}

The computation of the IO-contract quotient follows a similar reasoning. Given IO contracts $\cont = (I, O, \alist, \glist)$ and $\cont = (I', O', \alist', \glist')$, we want to compute $\cont / \cont' = (I_q, O_q, \alist_q, \glist_q)$ applying \eqref{eq:quotient}. First, the quotient is defined only if $I$ and $O'$ are disjoint, as the outputs of $O'$ cannot be inputs of the top-level. The inputs and outputs of the quotient are $I_q = (O' \setminus O) \setunion (I \setminus I')$ and $O_q = (O \setminus O') \setunion (I' \setminus I)$, respectively.
Setting once again $a = \land \alist$, $g = \land \glist$, $a' = \land \alist'$, and $g' = \land \glist'$,
the quotient assumptions and guarantees are given by \eqref{eq:quotient}.
As with composition, this expression is not a valid IO contract. As the quotient is defined as a maximum, we will refine the quotient operation in order to transform it into a valid IO contract.

The assumptions of the quotient are $a \land (\neg a' \lor g')$. Refining the quotient means to enlarge the assumptions---see \eqref{eq:refinement}. In order to respect the IO contract structure, we keep the assumptions $a$, and we add to them the guarantees $g'$ if $a \le a'$. From the resulting list we remove terms containing irrelevant variables (i.e., those not in $I_q$), thus generating a relaxation of the assumptions. The guarantees of the quotient are $(a' \land g) \lor \neg a  \lor (a' \land \neg g')$, an expression we refine using the following result:
\begin{theorem}\label{th:lmqefnq}
    Let $(a, g)$ and $(a',g')$ be contracts defined over a term algebra $T$. Let $\cont_q = (a_q, g_q)$ be the quotient $(a,g) / (a', g')$, as computed by \eqref{eq:quotient}. Suppose that $a'' \in T$ satisfies $a_q \le a''$ and $g'',g''' \in T$ satisfy $a' \land g' \land g'' \le a' \land g' \land g$ and $g''' \land a \le a' \land g'' \land a$. Then the contract $(a'', g''')$ is a refinement of $\cont_q$.
\end{theorem}
\ifAddProofs
\begin{proof}
    Since we are given $a_q \le a''$, we just have to verify the guarantees:
    $
    g''' \lor \neg a'' \le  g''' \lor \neg a_q =
    g''' \lor \neg a \lor (a' \land \neg g') =
    (g''' \land a) \lor \neg a \lor (a' \land \neg g')  \le 
    (a' \land g'' \land a) \lor \neg a \lor (a' \land \neg g') =
    (a' \land g'' \land g') \lor \neg a \lor (a' \land \neg g') \le 
    (a' \land g' \land g) \lor \neg a \lor (a' \land \neg g') =
    g_q \lor \neg a_q.
    $
    We conclude that $(a'', g''') \le \cont_q$.
\end{proof}
\fi
This theorem tells us how to obtain a correct refinement of the quotient.
Procedure \textproc{ContractQuotient} of Algorithm \ref{nlkqbl} shows we compute the quotient of IO contracts using Theorem \ref{th:lmqefnq} and the considerations above. 

\begin{example}
In Figure\ref{fg:comp-quotient}, we want to implement a system $M$ with contract
$(i, o', \allowbreak\{i \le 1\}, \{o' \le 2i + 1\})$ using a partial implementation $M'$ with contract $(i, o, \{i \le 2\}, \{o \le 2i\})$. We use \textproc{ContractQuotient} of Algorithm~\ref{nlkqbl} to find the specification $(I^q, O^q, \alist'', \glist''')$ of the missing component. From L\ref{qkgsdj}-L\ref{kbcg}, we obtain
$I^q = \{o\}$ and $O^q = \{o'\}$.
L\ref{qkgcqjhg}-L\ref{lqhlq} result in
$\alist'' = \textproc{RelaxWithContext}(\{i \le 1, o \le 2i\}, \emptyset, \{o\}) = \{o \le 2\}$.
L\ref{lkjxgsdb} yields
$\glist'' = \textproc{RefineWithContext}(\{o' \le 2i + 1\}, \{i \le 2, o \le 2i\}, \{o, o'\}) = \{o' \le o + 1\}$.
From L\ref{lkjdan}, we get
$\glist''' = \textproc{RefineWith-}\allowbreak\textproc{Context}(\allowbreak\{i \le 2, o' \le o + 1\}, \{i \le 1\}, \{o,o'\}) = \{o' \le o + 1\}$.
We obtain a specification only depending on the missing component's inputs and outputs.
\qeda
\end{example}

The computation of strong merging~\eqref{eq:merging} does not need refinements/relaxations and is given by the \textproc{ContractMerging} procedure of Algorithm \ref{nlkqbl}.
Finally, we consider contract refinement. We observe that a cumbersome issue with refinement is that \eqref{eq:refinement} requires the computation of complements. Complements would require us to carry out expensive expansions of the termlists. The following proposition allows us to compute the quotient without taking complements.
\begin{proposition}\label{poijzn}
Let $\cont = (a, g)$ and $\cont' = (a', g')$ be contracts over the term algebra $T$. Then $\cont \le \cont'$ if and only if $a' \le a$ and $g \land a' \le g' \land a'$.
\end{proposition}
\ifAddProofs
\begin{proof}
    Suppose that
    $g \land a' \le g' \land a'$. Then
    $g \lor \neg a' \le g' \lor \neg a'$.
    Since $\neg a \le \neg a'$, we have
    $g \lor \neg a \le g' \lor \neg a'$.
    Conversely, suppose that $g \lor \neg a \le g' \lor \neg a'$. Since $\neg a \le \neg a'$,
    $g \lor \neg a' \le g' \lor \neg a'$. Conjoining both sides with $a$ yields
    $g \land a' \le g' \land a'$.
\end{proof}
\fi

\vspace{-5mm}
\section{Case studies}
\label{sc:casestudies} 

We implemented \pacti, a tool that allows us to carry out system-level reasoning using assume-guarantee contracts.
\pacti supports all IO contract operations described in Section \ref{contractoperations}.
Its implementation of the contract algebra is orthogonal from that of the specification formalism in which contracts are expressed.
The first specification theory supported by \pacti is polyhedral constraints. 
This section presents the use of \pacti in several domains.

\subsection{Evaluating the end-to-end autonomy stack}

Evaluating and validating perception systems with respect to safety-critical requirements of an autonomous system is an active area of research~\cite{badithela2021leveraging,badithela2022evaluation,topan2022interaction}.
We explore contract-based design of an autonomy stack for a self-driving vehicle consisting of two components: an object detection module and a control module.
By imposing a system-level safety contract on the vehicle and by having available the contract of the control module, obtained by \emph{merging} contracts of control submodules, we use the \emph{quotient} to obtain a specification for the object detection module. 
Our model of the perception subsystem takes as input the distance to the object that must be recognized and outputs the true positive rate at which the object is detected. The controller's model takes as input the true positive rates of the perception component and outputs the probabilities that certain safety properties will be satisfied.
We want to characterize lower bounds on the perception component's true positive rates---implementations of the perception model that satisfy these requirements will enable the end-to-end stack to satisfy the system-level satisfaction probability lower-bound.

\begin{figure}[t]
    \centering
    \includegraphics[width=\textwidth]{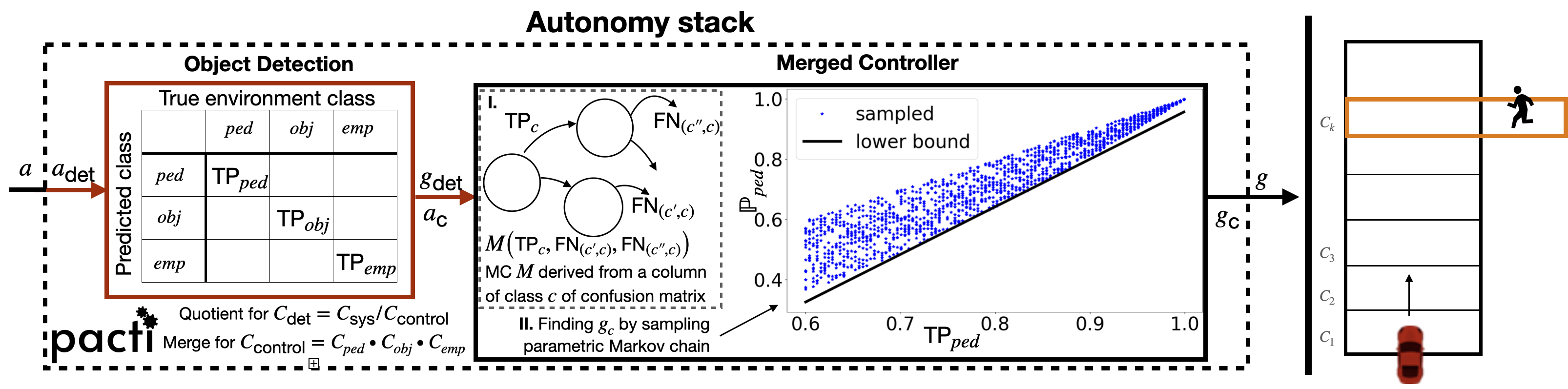}
    \vspace{-20pt}
    \caption{\footnotesize
    An autonomous vehicle must respect certain safety constraints with respect to what is present on the crosswalk: a pedestrian, another object, or nothing.
    The autonomy stack consists of two elements: a perception module and a controller made of three sub-controllers, one for each class that can be recognized by the perception component.
    Given the system-level contract of the autonomy stack \(\cont_{\text{sys}} = (a,g)\) and the control contract \(\cont_{\text{control}} = (a_{\text{control}},g_{\text{control}})\), the quotient operator in \pacti allows us to compute the contract for the object-detection module, \(\cont_{\text{det}} = (a_{det},g_{det})\).
    }
    \label{fig:autonsys}
\end{figure}
\paragraph{Example and system contract.} Suppose that we want our vehicle to stop at a crosswalk only if it detects a pedestrian.
The perception task is to correctly detect the object at the crosswalk.
This object can belong to three classes: pedestrian (denoted \emph{ped}), an object that is not a pedestrian (\emph{obj}), and the background or empty class (\emph{emp}). 
The test set evaluations of a learned object detection model can be represented by a confusion matrix, where the \((i,j)^{th}\) element represents the probability with which the model classifies an object with true label \(j\) to be of class \(i\).
The vehicle operates according to three controllers: when one of the three object classes is detected, the corresponding controller engages.
Each of these controllers is synthesized to satisfy a safety property given in linear temporal logic---we use the vehicle dynamics and safety properties considered in~\cite{badithela2021leveraging}.
Since object detection is not perfect, the controller can only satisfy its safety specification probabilistically. Thus, we use probabilistic model checking to compute 
the satisfaction of the specification. 

At any time step \(t\), the vehicle chooses its action (the controller) according to the control strategy corresponding to the object detected at time \(t\).
We want the system contract to be \(\cont_{\text{sys}} = (d_l \leq d \leq d_u, g_{{ped}} \land g_{{obj}} \land g_{emp})\), where \(g_{c} := m_c d + b_c \leq \mathbb{P}(s_0 \models \varphi_c) \), for reals \(m_c\) and \(b_c\), and \(c \in \{ped, obj, emp\}\). The guarantees \(g_{c}\) represent lower bounds on the satisfaction probability of \(\varphi_c\) as a function of the distance \(d\) to the object to be detected. The formulas \(\varphi_c\) correspond to formal requirements on the car when the object of the crosswalk is of class \(c\). For brevity, we use \(\mathbb{P}_c\) to denote \(\mathbb{P}_c(s_0 \models \varphi_{c})\).
This top-level contract assumes that the distance to the crosswalk is bounded by $d_l$ and $d_u$. It guarantees that the probabilities that the safety properties will be satisfied are bounded below by an affine function of $d$. Figure~\ref{fig:autonsys} illustrates this example. 


\paragraph{Controller contract.} The contract for each of the three controllers takes as inputs the true positive rates of the object classes and outputs the probability that the vehicle will satisfy a certain safety property when operating under said controller.
Following~\cite{badithela2021leveraging},
we compute the probability that the controller satisfies the safety property $\varphi_c$
by constructing a Markov chain whose transition probabilities are derived from the confusion matrix of the perception component.
For the car-pedestrian example, the probability of satisfying \(\varphi_c\) depends mainly on the true positive rate of the class \(c\). Therefore, the affine lower bounds on satisfaction probability that the controller can guarantee are of the form \(m_l(\texttt{TP}_c) + b_l\leq \mathbb{P}_c \),
where \(m_l\) and \(b_l\)
are reals. 
We determine this affine bound by fixing $\texttt{TP}_c$ and computing the probability that the controller will satisfy its property $\varphi_c$ for several values of false negatives for class \(c\) (with $c' \ne c$).
Note that we determine these affine bounds only for the range of the true positives that we sample over.
We end up with the following contract for the controller corresponding to object \emph{c}:
$
\cont_{\text{c}} = (lb \leq \texttt{TP}_c ,\; a_l(\texttt{TP}_c) + b_l\leq \mathbb{P}_c 
)
$.
After computing these three contracts, we merge them to get the overall control contract: \(\cont_{\text{control}} = \cont_{ped} \bullet \cont_{obj} \bullet \cont_{emp}\).

\paragraph{Object detection contract.} Using the system-level contract $\cont_{\text{sys}}$ and the controller contracts, we can compute the specification of the perception subsystem through the quotient \(\cont_{\text{det}} = \cont_{\text{sys}}/\cont_{\text{control}}\). The resulting contract imposes bounds on the true positive rates of each object class. We illustrate the results numerically here for an instantiation of this car-pedestrian example. The system contract is set to
$\cont_{sys} = (1 \leq d \leq 10, 0.99(1 - 0.1d) \leq  \mathbb{P}_{ped} \land  0.8(1 - 0.1d) \leq \mathbb{P}_{obj}  \land 0.95(1 - 0.1d) \leq \mathbb{P}^{emp})$, i.e., we assume the distance to the crosswalk is bounded between 1 and 10, and the contract guarantees that the probabilities of satisfaction of the three properties are bounded below by affine functions of the distance.
With lower bounds derived by solving a linear program, the controller contracts are computed to be 
$\cont_{ped} = (0.6 \leq \texttt{TP}_{ped}, 1.58\texttt{TP}_{ped} - 0.622 \leq \mathbb{P}_{ped})$,
$\cont_{obj} = (0.3 \leq \texttt{TP}_{obj}, 0.068\texttt{TP}_{obj} + 0.93 \leq \mathbb{P}_{obj})$, and 
$\cont_{emp} = (0.6 \leq \texttt{TP}_{emp}, 0.2\texttt{TP}_{emp} + 0.799 \leq \mathbb{P}^{emp})$. That is, they impose an affine lower bound on probabilities of formula satisfaction using the true detection rates.
The resulting quotient leads to an object detection contract with true positive rates lower bounded by affine functions of the distance \(d\):
$
\cont_{\text{det}} = \big(  1 \leq d \leq 10,\; (1.02 - 0.063d \leq \texttt{TP}_{ped}) \land (0.6 \leq \texttt{TP}_{ped}) \land  
(0.3 \leq \texttt{TP}_{obj}) \land  (0.75-0.474d \leq \texttt{TP}_{emp}) \land (0.6 \leq \texttt{TP}_{emp})\big)
$.

We showed how the contract operators implemented in \pacti are useful in characterizing bounds of the confusion matrix to achieve the desired probability of satisfaction for a system-level temporal logic formula. This is useful in providing specifications to designers responsible for object detection. Thus, by evaluating an object detection model on a test set and checking if the resulting confusion matrix satisfies the requirements on true positive bounds as in $\cont_{\text{det}}$, we can conclude whether the end-to-end autonomy stack with controller designed according to \(\cont_{\text{control}}\) will satisfy the system-level requirements.

\subsection{Trajectory planning for multi-agent systems}
This case study illustrates the use of the \emph{merge} operator in \pacti to solve a multi-agent navigation application on a grid world. Multi-agent path-finding (MAPF) is the problem of finding paths that multiple agents can concurrently follow to their target locations without colliding with each other. 
Finding a path for a single agent is efficiently solved using the A* algorithm~\cite{hart1968formal}, but when multiple agents are introduced the problem increases in complexity. MAPF is NP-hard, as the state space grows exponentially with the number of agents~\cite{surynek2010optimization,yu2013structure}.
This problem has been studied for various applications, including warehouse robots~\cite{wurman2008coordinating}, traffic control~\cite{dresner2008multiagent}, aviation~\cite{pallottino2007decentralized}, and video games~\cite{ma2017feasibility,silver2005cooperative}.
\begin{figure}[t]
    \begin{minipage}{.50\textwidth}
    \includegraphics[width=\linewidth,trim={0.0cm 0.0cm -0.0cm 0.0cm}]{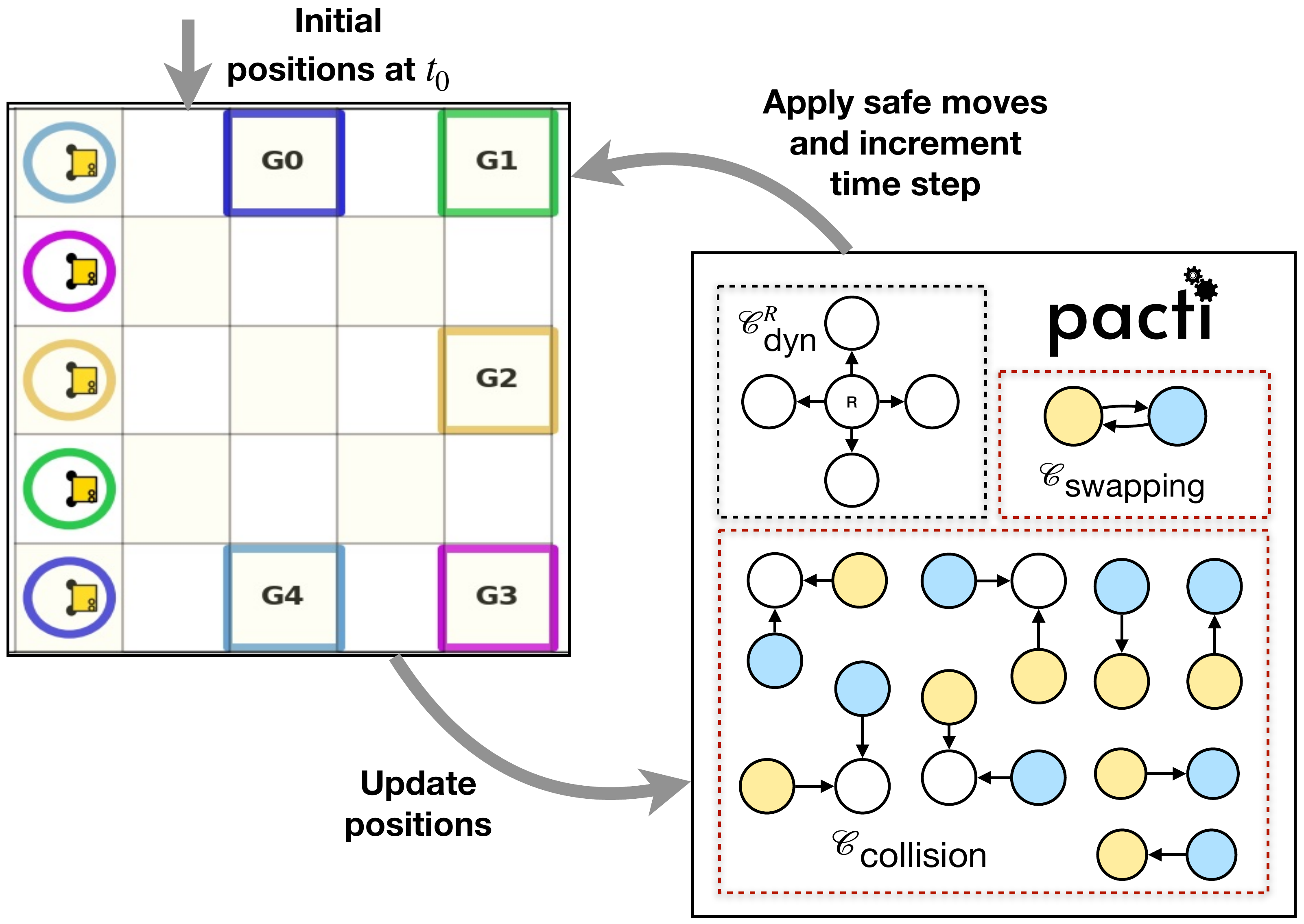}
\subcaption{\label{fig:multiagent-blocks}Overview of the process per time step}
\end{minipage}
\hspace{1mm}
     \begin{minipage}{.35\textwidth}
\includegraphics[width=\linewidth,trim={0.0cm 0.0cm -0.0cm 0.0cm}]{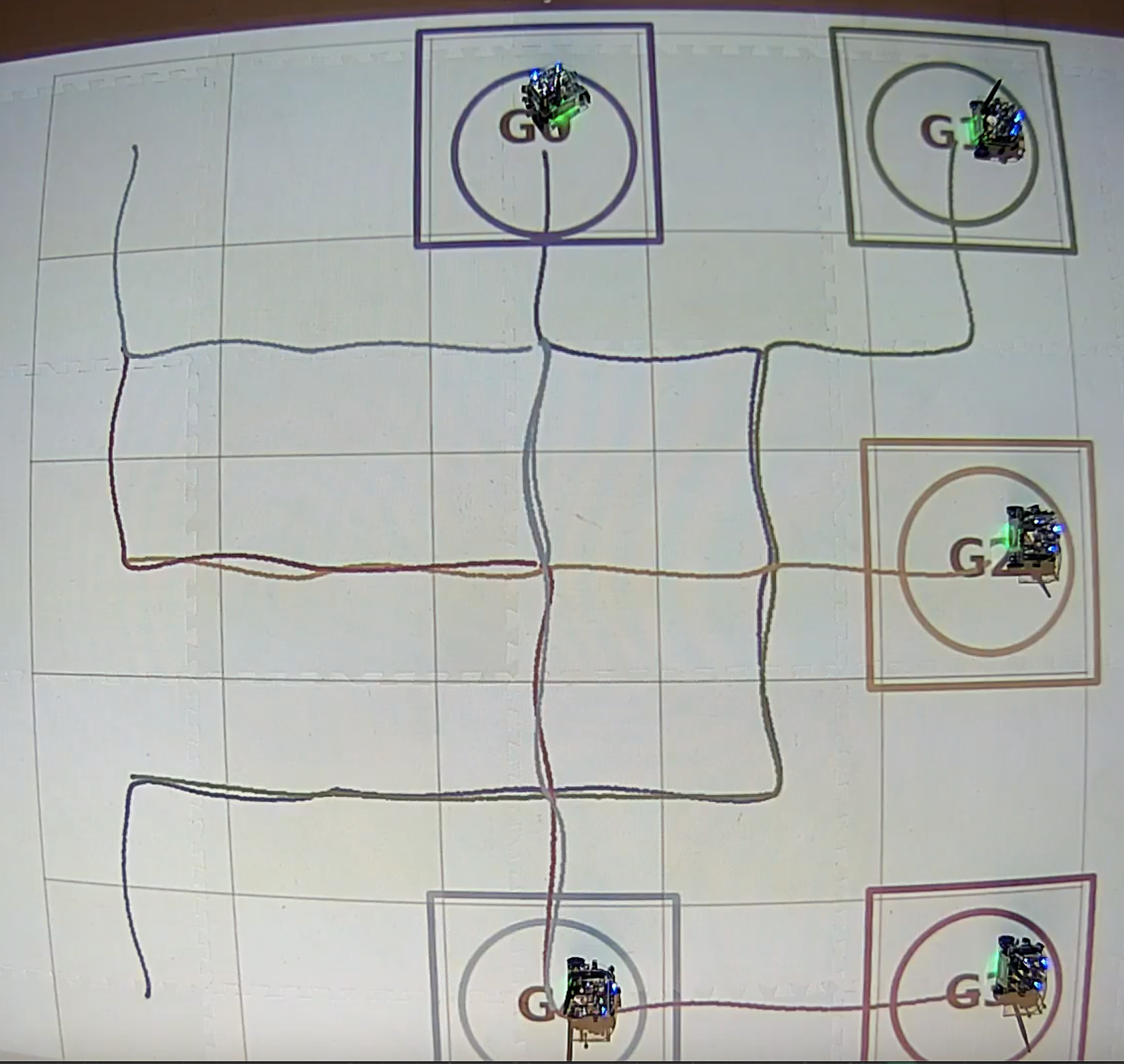}
\subcaption{\label{fig:multiagent-result}Resulting strategy}
  \end{minipage}
  \caption{\footnotesize
    MAPF problem framed using contracts in \pacti. (a)~Each time step is treated as a viewpoint, and \pacti is used to generate a list of safe moves, from which the next move is selected and applied. This process repeats until the goal positions are reached. (b)~The conflict-free trajectories executed on the Robotarium using \pacti.}
    \label{fig:multiagent-fig}
\end{figure}

Consider five robots in a grid world. Each robot starts at its initial position and needs to reach its goal position without colliding with other robots. The dynamics of the robots allow horizontal or vertical transitions to the adjacent neighboring cells or waiting in the same cell. We model this problem in discrete time, with all robots simultaneously executing their next moves.
The layout of the grid world is shown in Figure~\ref{fig:multiagent-fig}. In this problem we will ensure that there are no vertex conflicts---when multiple agents share a single cell---and no swapping conflicts---when two agents swap positions. For an overview of the different objectives and conflict types, see~\cite{stern2019multi}.

At each discrete time $t$, we model the possible movements of a robot via a dynamics contract. This contract guarantees that the location of the robot in the next step is reachable from the current position of the robot. That is, 
we encode the dynamics of each robot $r$ in the contract $\cont_{\text{dyn}}^r = \big(a_{\text{dyn}}^r, g_{\text{dyn}}^r \big)$
with the assumptions $a_{\text{dyn}}^r$ and guarantees $g_{\text{dyn}}^r$ defined as
$
a_{\text{dyn}}^r = \:
(t_i = t) \: \land 
(x^r_i = x^r) \:\land \: (y^r_i = y^r),
$ and
$
%
g_{\text{dyn}}^r =    
(t_{i+1} = t_i + 1) \land 
(|x^r_{i+1} - x^r_i| + |y^r_{i+1} - y^r_i| \leq 1)
$
with $x^r$ and $y^r$ denoting the current coordinates of robot $r$ and $i$ is the current time step. This contract assumes that the robot is at the given position and guarantees that it will move at maximum one cell horizontally or vertically.

After this, we merge the dynamics contracts for all robots at the current time step to create the contract describing the dynamics for all robots on the grid $\cont_{\text{dyn}}$, which describes all possible moves the robots can take, but does not take into account possible conflicts.
To prevent vertex conflicts, we create contracts whose guarantees enforce that the distance between each robot pair is equal to or larger than one cell, and thus ensure no cell will be occupied by more than one robot. This constraint is given as $g_{\text{collision}}= |x^{r_1}_{i+1}-x^{r_2}_{i+1}| + |y^{r_1}_{i+1}-y^{r_2}_{i+1}| \geq 1$. As \pacti only allows linear inequalities to describe the requirements, we define four separate collision constraints for each pair of robots. In Figure~\ref{fig:multiagent-blocks} possible collisions between two robots are illustrated under $\cont_{\text{collision}}$---on the left the four collision configurations are depicted and on the right we show special cases of each where one robot directly crashes into the other.
These four constraints are mutually exclusive, but as the robots can only be in one configuration at a time, it allows us to consider them individually to find all possible move options. 
For the robot pair $r_1,r_2$ we model the collision contracts as
$\cont_{\text{collision}}^i = \big( a_d,\quad x^{r_1}_{i+1} + (-1)^{\lfloor (i+1)/2 \rfloor} x^{r_2}_{i+1} + (-1)^{i+1} y^{r_1}_{i+1} + (-1)^i y^{r_2}_{i+1} \leq -1  \big)$ for $1 \le i \le 4$ and
$a_d := \text{distance}(r_1,r_2) \geq 1$, i.e., assuming no collision at the current time step.

Next we define the contract that ensures that the robots will stay collision free when transitioning to another cell, preventing swapping conflicts. We manipulate the $x$ and $y$ coordinates of the two robots, $r_1$ and $r_2$, in the pair as
 $\delta_x = (x^{r_1}_0 - x^{r_2}_0) * (x^{r_1}_1 - x^{r_2}_1)$ and
 $\delta_y = (y^{r_1}_0 - y^{r_2}_0) * (y^{r_1}_1 - y^{r_2}_1)$.
Now we can define the contract using $\delta_x$, $\delta_y$, and $\text{distance}(r_1,r_2)$ as inputs
$\cont_{\text{swapping}} = (a_d, \quad \delta_x + \delta_y \geq 1)$.

We find the move candidates by merging the dynamics contract $\cont_{\text{dyn}}$ separately with each of the collision contracts $\cont_{\text{collision}}^i$ for each pair of robots.
Possible next moves that satisfy one of these four contracts ensure that the corresponding robot pair is safe with respect to each other.
To ensure that all robots remain collision-free, we only allow move candidates that are safe for all robot pairs.

Lastly, we check the swapping contract $\cont_{\text{swapping}}$ and discard any move candidates that do not satisfy the contract. Any remaining move candidate is now guaranteed to be a safe move with respect to the robot's dynamics, vertex conflicts, and swapping conflicts. 
Having obtained a set of valid moves,
we randomly choose a move with the goal of reducing the euclidean distance to the goal position for each robot. 
In case the robots cannot improve their position, a random move is taken. The robots will each take the chosen move. The entire process is repeated for the next time step until all robots have reached their goals.

Using \pacti in this application allows us to modularly add contracts for additional robots and other constraints to the problem. We showed how \pacti can be used to model a dynamical system by assigning contracts to the dynamics and valid positions at each decision step.
We used Robotarium, a remote-access, multi-robot testbed~\cite{wilson2020robotarium}, to implement the resulting strategy for multiple robots on hardware shown in Fig.~\ref{fig:multiagent-result}. 

\subsection{Specification-based synthetic biology}

\begin{figure}[!tb]
    \centering
    \includegraphics[width=0.9\textwidth]{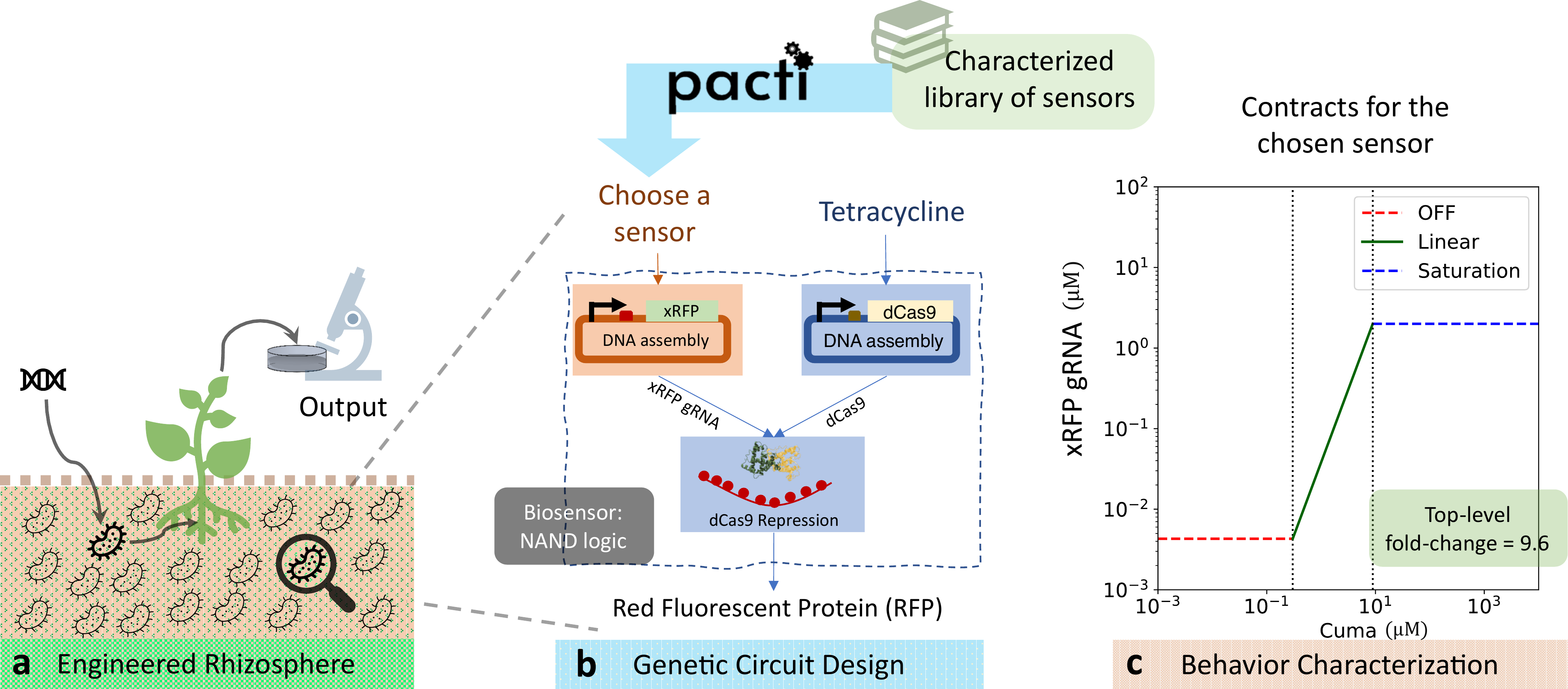}
    \caption{\footnotesize \textbf{Specification-based synthetic biology using \pacti}. (a) Rhizosphere is the region near the plant roots that is affected by the soil's microorganisms. The goal with the engineered rhizopshere is to measure the plant protein expression in response to a target gene introduced through engineered bacterial cells. (b) The bacterial cell is engineered as a biosensor that detects two input signals and represses the free production of its output only when both inputs are present. (c) Using \pacti, we can choose a sensor from a characterized library to achieve the highest fold-change for the NAND logic gate. Three contracts that model the specification of the chosen sensor are shown.}
    \label{fig:synbio}
\end{figure}

This case study shows the use of \pacti to aid the design of engineered biological circuits for altering and observing the dynamics of bacterial gene transfer in soil and its effects on protein expression in plants (schematic shown in Figure~\ref{fig:synbio}a). As the full system design is large, we focus on the design of a biological NAND gate system. This system has three subsystems, as shown in Figure~\ref{fig:synbio}b. We will use contract \emph{composition} as part of an optimization strategy to maximize the system fold-change, defined as the ratio between the on and off levels of the system's output. The \emph{quotient} will be used to infer a specification of a missing subsystem that needs to be designed in our system. We say that an input or an output is ``ON'' when its level is higher than a minimum  threshold, and is ``OFF'' when its level is lower than a maximum threshold. We formalize these notions with polyhedral constraints in the contract descriptions. 

The biological NAND gate consists of the three subsystems:
1) a sensor with tetracycline (aTc) input that outputs a dCas9 protein, 2) another sensor (to be chosen) that outputs the xRFP-gRNA protein, and 3) a dCas9 repressor subsystem, which takes as inputs the xRFP-gRNA and the dCas9 proteins. When both inputs to this repressor subsystem are ON, it suppresses the output, RFP, a red fluorescent protein. RFP is also the output of the top-level system. RFP is ON only when either of the sensors inputs are OFF. In this way, the system behavior is that of a NAND logic gate. We will denote the contract for the tetracycline sensor as $\cont_{\text{aTc}}$, and the contract for dCas9 repression mechanism as $\cont_{\text{dCas9}}$. Our first task will be to find the second sensor (choose spec for subsystem 2) that maximizes the top-level system fold-change. Then we will consider the task of finding a specification for the repression component (subsystem 3) assuming that the contracts for both sensors and the top-level system are available.

\paragraph{Modeling the specifications and constructing a library of parts.}
First we represent a set of sensors using contracts.
We build a library of sensor contracts using the experimental data for the sensors in the \textit{Marionette} bacterial cell strain~\cite{marionette}. 
Each sensor has three characteristic behaviors: 1) the off state, where the output of the sensor stays close to zero (the non-zero expression in this state is termed as ``leaky response''), 2) the linear rate of output, where the sensor output responds linearly to the input (in log scale), and 3) the saturation state, where the sensor output saturates to a maximum constant value. The three contracts for a sensor, $s$, with input $u$ and output $y$ are
$\cont^{\text{off}}_s = (u \le u_{\text{start}}, y \le y_{\text{leak}})$,
$\cont^{\text{sat}}_s = (u \ge u_{\text{K}}, y \ge y_{\text{max}})$, and
$\cont^{\text{lin}}_s = (u_{\text{start}} \leq u \leq u_{\text{K}}, y = m u + b)$,
where $m$ and $b$ are constants. The subscripts ``start'' and ``K'' denote the input threshold values of the end of the OFF behavior and the linear regime, respectively, whereas ``leak'' and ``max'' denote the leaky and the maximum values of the output. The ``off'', ``lin'', and ``sat'' superscripts represent the off, linear, and saturated viewpoints. Figure~\ref{fig:synbio}c shows the three contracts of one of the sensors constructed from the experimental data.

\paragraph{Contract composition \& sensor selection.}
The library of sensors we built contains data for 14 sensors. One of these sensors detects aTc, which we choose for our subsystem 1. The contract for aTc, $\cont_{\text{aTc}}$, can be constructed in three viewpoints as described above by using $K = 0.013, y_{\text{max}} = 1, \text{start} = 0.0018, y_{\text{leak}} = 4.9 \times 10^{-3}, m = 88.84, b = -0.15$. For subsystem 3, we write the dCas9 repression mechanism contract as $\cont_{\text{dCas9}}^{\text{lin}} = (0.3 \leq \text{xRFP} \leq 1 \land 0.1 \leq \text{dCas9} \leq 0.8,\:\: \text{RFP} + 2\:\text{xRFP} + 0.1\:\text{dCas9} \leq 5)$, with similar contracts for the off and saturated viewpoints.

To select the sensor for subsystem 2, we evaluate the performance of the system that is obtained when this sensor is chosen from the library of remaining 13 sensors.
This problem is described in Figure~\ref{fig:synbio}b.
For each sensor in the library with contract $\cont_{\text{s}_i}$, we use \pacti to compute the system-level contract by composing the chosen sensor contract, with the available subsystem contracts: $\cont_{\text{sys}} = \cont_{\text{aTc}} \parallel \cont_{\text{s}_i} \parallel \cont_{\text{dCas9}}$. When computing this composition for some of the sensors in the library, \pacti returns the error ``unsatisfiable in the given context.'' This means that the guarantees of this sensor are insufficient to meet the assumptions of the component to which it drives outputs (the dCas9 repressor subsystem). Thus, \pacti allows us to identify potential design errors.
In order to choose a sensor among those that yield valid compositions, we use the fold change of the system $F = \text{RFP}_{\text{on}}/\text{RFP}_{\text{off}}$ as a performance criterion and select the sensor that maximizes this number.
The final chosen sensor that achieves the highest fold-change for the NAND gate is the ``Cuma'' sensor shown in Figure~\ref{fig:synbio}c. 

\paragraph{Contract quotient to find the specifications of missing parts.}
We now consider a different problem. Suppose that we have chosen the two sensors (subsystems 1 and 2), and we are also given a desired top-level system contract $\cont_{\text{sys}}$ that the system must meet. We use the quotient in \pacti to find the specification of the missing object: the dCas9 repression mechanism (subsystem 3). For example, we assume that we have the sensors ``Sal'' and ``aTc'' for which we have $\cont_{\text{Sal}}^{\text{lin}} = (0.9 \leq \text{Sal} \leq 43.0, \: \: 0.03\: \text{Sal} - \text{xRFP}  + 0.02 = 0)$ and $\cont_{\text{aTc}}^{\text{lin}} = (0.0018 \leq \text{aTc} \leq 0.013,\:\:88.84\:\text{aTc} - \text{dCas9} + 0.15 = 0)$. For the top-level system, we have $\cont_{\text{sys}}^{\text{lin}} = (0.909 \leq \text{Sal} \leq 42.57 \land 0.0018 \leq \text{aTc} \leq 0.012, \text{RFP} + \text{Sal} + \text{aTc} \leq 1.29)$. Using the quotient, we obtain $\cont_{\text{dCas9}}^{\text{lin}} = \cont_{\text{sys}}^{\text{lin}} / (\cont_{\text{Sal}}^{\text{lin}} \parallel \cont_{\text{aTc}}^{\text{lin}}) = (0.05 \leq \text{xRFP} \leq 1.33 \land 0.31 \leq \text{dCas9} \leq 1.29, \text{RFP} + 0.01\:\text{dCas9} + 32.5\:\text{xRFP}\leq 1.29)$ as the contract for the dCas9 mechanism when it represses the RFP level. For brevity, we have only shown the condition when both sensors are switched on in the linear regime. The resulting contract for subsystem 3 guarantees that it represses the RFP level dependent on its inputs, xRFP and dCas9. We can provide this missing-component contract to an expert for independent implementation.


\subsection{Signal processing pipelines in digital integrated circuits}
\begin{figure}[t!]
    \centering
    \includegraphics[width=.7\linewidth]{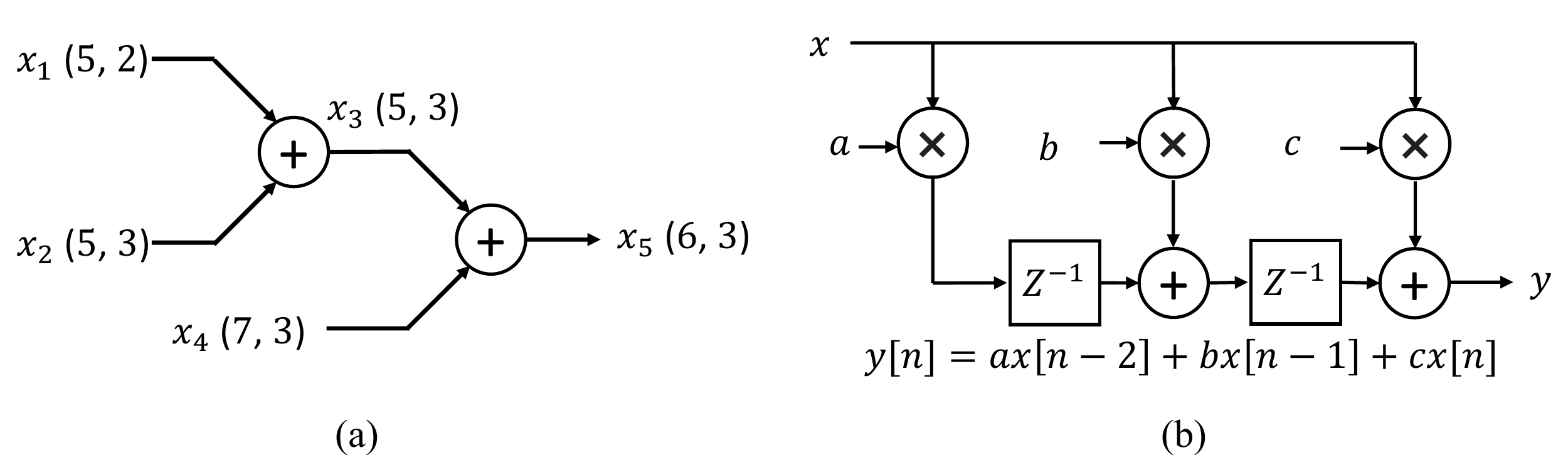}
    \vspace{-2mm}
    \caption{\footnotesize (a) An example of two adders. The word-length is denoted beside the name of the number. (b) An example of a 3-tap digital filter.}
    \label{fig:dsp-example}
\end{figure} 
The numerical representations of digital signal processing algorithms must often be translated from floating-point to fixed-point when the algorithms are implemented in hardware.
To reduce implementation costs and increase performance,
one of the objectives of this translation is to use the smallest possible fixed-point representations that allow the algorithm to operate within acceptable error bounds.
This is a time-consuming and error-prone step that relies on statistical quantities gathered from simulations~\cite{Cantin02,Constantinides03,Keding98,Kim98,Shi04,Sung95}.
As simulation yields limited coverage,
techniques for verifying digital signal processing design have been proposed in the last decades~\cite{Cox12,Fang03,Simic21}.

This case study demonstrates the use of \pacti to find error bounds of fixed-point digital signal processing algorithms
and perform local word-length optimization efficiently through contract \emph{composition}, \emph{quotient}, and \emph{refinement}.
We have two use cases. The first computes the bound in arithmetic error of a system obtained by composing several arithmetic operations, each introducing its own fixed-point error.
The second imposes a total error budget on the system and uses the quotient to find the maximum error that one of the components may have while meeting the system-level budget. This operation is helpful to size circuits.
For simplicity, all numbers in this case study are unsigned.

\smallskip
\paragraph{Contract formulation of fixed-point operations.}
We first formulate fixed-point numbers and operations as contracts.
The word-length of a fixed-point variable $x$ is defined as a tuple $(x_n, x_p)$.
$x_n$ denotes the number of bits to encode the fixed-point number, while $x_p$ is the number of bits to encode the integer part~\cite{Constantinides03}.
Accordingly, the fractional part is encoded in $x_n - x_p$ bits.

We model a fixed-point number $x$ using two variables $x_e$ and $x_a$ to represent the relationship between inputs and outputs.
$x_a$ is the maximum value that the variable can take, and $x_e$ is the maximum error between the value of the fixed-point number and the ideal value of $x$.
When $x$ is a constant coefficient, $x_e$ is the quantization error of the coefficient.

\begin{enumerate}
    \item \textit {General Operation.}
    Given two input numbers $x$ and $y$ and an output number $z$ with word-lengths $(x_n, x_p)$, $(y_n, y_p)$, and $(z_n, z_p)$, respectively,
    we form the contract for operation $z^* = f(x^*, y^*)$ as $\cont_{op} = (a_{op}, g_{op})$ that encodes the relation between variables $x_e, x_a, y_e, y_a, z_e, z_a$ as follows:
    {\footnotesize
    \begin{equation}\label{eq:contract-op}
    \begin{aligned}
        a_{op} &= \max_{\substack{ 0 \leq x \leq x_a \\
                              0 \leq y \leq y_a}} f(x, y) < 2^{z_p} \\
        g_{op} &=  ( z_e \leq C)  \land 
        (z_a \leq 2^{z_p} - 2^{z_p-z_n}) \land
        \left(z_a \leq \max_{\substack{ 0 \leq x \leq x_a,
                              0 \leq y \leq y_a}} f(x, y)\right),
    \end{aligned}
    \end{equation}}%
    where $C = \max_{\substack{ 0 \leq x \leq x_a,
    0 \leq y \leq y_a}} (f(x, y) - f(x - x_e, y - y_e))  + \max(0, 2^{z_p}(2^{-z_n} - 2^{-(w_n - w_p + z_p)}))$ and 
    $(w_n, w_p)$ is the ideal minimum word-length that ensures no truncation loss and overflow hazards~\cite{Constantinides03}.

    The assumption ensures no overflow occurs by checking that the maximal possible result can be fit into the range of the output variable.
    The guarantee is a conjunction of three clauses.
    The first clause bounds the output error considering the errors propagating from the input and the truncation error.
    The second clause states that the maximum output value is bounded by its fixed-point number representation.
    The third clause bounds the output value using the bounds on the inputs.
    Once we determine $w_n$, $w_p$, and the maximization terms for an operation, we can form the contract for it, even without the implementation details of the operation.

    \item \textit{Addition.}
    Using~\eqref{eq:contract-op}, we can derive the contract 
    $\cont_{add}$ for addition as follows:
    {\footnotesize
    \begin{equation}\label{eq:contract-add}
        \begin{aligned}
        a_{add} &= x_a + y_a < 2^{z_p} \\
        g_{add} &=  (z_e \leq x_e + y_e + C)  \land (z_a \leq 2^{z_p} - 2^{z_p-z_n} \land
        (z_a \leq x_a + y_a) ),
    \end{aligned}
    \end{equation}}%
    where $C = 2^{z_p}(2^{-z_n} - 2^{y_p - \max{(x_n, y_n - y_p + x_p)} - \min{(y_p-x_p, x_p - y_p)} - z_p}$.
    Note that the contract for addition includes only polyhedral constraints.

    \item \textit{Multiplication.}
    The contract $\cont_{mult} = (a_{mult}, g_{mult})$ for multiplication can be defined from~\eqref{eq:contract-op} as follows:
    $a_{mult} = x_a y_a < 2^{z_p}$ and
    $g_{mult} =  (z_e \leq y_a x_e + x_a y_e - x_e y_e + 2^{z_p}(2^{-z_n} - 2^{-(x_n + y_n - x_p - y_p + z_p)}))  \land 
    (z_a \leq 2^{z_p} - 2^{z_p-z_n}) \land
    (z_a \leq x_a y_a)$.
    
    This contract includes non-polyhedral constraints due to the multiplications of the variables in the guarantee.
    However, if one of the inputs to the multiplication is a constant, the resulting contracts involve only polyhedral constraints.
    The operation of multiplying an input with a constant coefficient is common in signal processing.
\end{enumerate}

\smallskip
\paragraph{Applying \pacti in the verification and optimization of fixed-point operations.}
We consider how we can use \pacti to reason about fixed-point error specifications
and to generate meaningful results both for designers and optimization tools.
Consider the two fixed-point adders shown in Figure~\ref{fig:dsp-example}a. 
We encode the contracts of the two adders using the formulation in ~\eqref{eq:contract-add} as $\cont^1$ and $\cont^2$.
Then we encode three contracts $\cont^{x_1}, \cont^{x_2}$, and $\cont^{x_4}$ to represent the constraints on the three inputs $x_1$, $x_2$, and $x_4$ based on the word-lengths and assuming that the input has no errors.
For example, the contract $\cont^{x_1}$ for the input $x_1$ is formulated as $(\text{True}, 0 \leq {x_1}_a \leq 7.75 \land {x_1}_e = 0)$.

To obtain the error specification for the entire system,
we compute the composition $\cont^{x_1} \parallel \cont^{x_2} \parallel \cont^{x_4} \parallel \cont^1 \parallel \cont^2$ using \pacti. The tool gives an error indicating that a system cannot be built because
the guarantees of the first adder are insufficient to satisfy the assumptions of the second adder.
This error means that an overflow might occur in $x_3$.
Indeed, when $x_1 = 11.111$ and $x_2 = 111.11$, the result $1011.101$ would cause an overflow in $x_3$, which only has three bits for the integer part.
If we further limit the maximum value of the inputs using the constraints $x_1 \le 2$, $x_2 \le 3.75$, and $x_4 \le 0.03125$, we obtain the system-level contract $(\text{True},g = {x_5}_a \leq 5.8125 \land{x_5}_e \leq 0.1875)$. 
The contract gives bounds for the value and the maximum error of the system-level output.

We consider a different situation.
In the system of Figure~\ref{fig:dsp-example}a, suppose that we have a top-level contract $\cont_{sys}$, requiring that the resulting system has an output error smaller than $0.1$, and we have the same input constraints as before. Our objective is to find the word-length of $x_3$ that would allow the system to meet its objective.
First, we compute the quotient $\cont_{local} = \cont_{sys} / (\cont^{x_1} \parallel \cont^{x_2} \parallel \cont^{x_4})$ to get the local specification for components affected by $x_3$.
From this specification, we update the word-length of $x_3$, compute the corresponding local contracts $\cont^{1} \parallel \cont^{2}$ for the two adders (both contracts depend on $x_3$), and check if the local specification is satisfied using refinement $(\cont^{1} \parallel \cont^{2}) \le \cont_{local}$.
This procedure continues until we find a word-length that satisfies the refinement relation.
This way we locally optimize the word-length of $x_3$. The result is that ${x_3}_n = 6$.

As a final example, consider the system shown in Figure~\ref{fig:dsp-example}b. By setting the coefficients to $a = 0.2$, $b = 0.6$, and $c = 0.2$, we obtain a weighted moving average filter. 
Using a similar approach as described, we compute the system-level error from subsystem specifications.
Our contract-based methodology yields a maximum error of $0.769$, while
the enumeration of all input combinations yields $0.688$.
Our obtained bound is pessimistic because each contract considers the worst-case scenario, which might not occur at the same time. On the other hand, its computation is vastly more tractable.

These examples illustrate that
we can use contract operations to obtain upper bounds for variable errors without enumerating all input combinations, which is crucial for performing optimization with many iterations.
In other words, \pacti can leverage contract-based design for combining formal methods with optimization
to reason about fixed-point representations in digital signal processing system design.

\vspace{-2mm}
\section{Discussion and concluding remarks}
\vspace{-2mm}
\vspace{-3pt}

We presented \pacti, a method to ease the broader use of assume-guarantee reasoning for system analysis and design.
Currently, \pacti supports the verification of refinement and the operations of composition, quotient, and strong merging. The program supports specifications written as polyhedral constraints.
Our objective is to develop a platform that can grow over time as more features are implemented.
Notably absent from the current implementation are support for the
verification of contract satisfaction by a component, synthesis of run-time monitors from contracts, and hypercontracts.
Moreover, we have only implemented support for polyhedral constraints. We plan to support other popular constraint formalisms, such as LTL and nonlinear constraints.
In addition, we plan to handle other contract operations either directly through implementations or through methodology.

\bibliographystyle{style/lncs/splncs04}
\bibliography{support/references}



\end{document}